\documentclass[5p,twocolumn,times,number]{elsarticle}

\usepackage{graphicx}
\usepackage{amsmath}   
\usepackage{caption}
\usepackage{subcaption}
\usepackage{lineno}
\usepackage{isomath}

\newcommand{\fig}[1]{Fig.~\ref{#1}}
\newcommand{\GeV}[1]{#1~GeV}

\begin{document}

\begin{frontmatter}

\title{A Highly Selective First-Level Muon Trigger With MDT Chamber Data\\ for ATLAS at HL-LHC}

\author{H.~Kroha}
\author{S.~Nowak\corref{cor}}
\ead{nowak@mpp.mpg.de}
\author{on behalf of the ATLAS Muon Collaboration}

\cortext[cor]{Corresponding author}

\address{Max-Planck-Institut f\"ur Physik, Munich, Germany}

\begin{abstract}
Highly selective triggers are essential for the physics programme of the ATLAS experiment at HL-LHC where the instantaneous luminosity will be about an order of magnitude larger than the LHC instantaneous luminosity in Run 1.
The first level muon trigger rate is dominated by low momentum muons below the nominal trigger threshold due to the moderate momentum resolution of the Resistive Plate and Thin Gap trigger chambers.
The resulting high trigger rates at HL-LHC can be sufficiently reduced by using the data of the precision Muon Drift Tube chambers for the trigger decision.
This requires the implementation of a fast MDT read-out chain and of a  fast MDT track reconstruction algorithm with a latency of at most 6~$\mu$s.
A hardware demonstrator of the fast read-out chain has been successfully tested at the HL-LHC operating conditions at the CERN Gamma Irradiation Facility.
The fast track reconstruction algorithm has been implemented on a fast trigger processor.

\end{abstract}


\end{frontmatter}

\section{Motivation}

The first level muon trigger of the ATLAS experiment~\cite{ATLAS2008} is based on Resistive Plate Chambers (RPC) in the barrel region and Thin Gap Chambers (TGC) in the end-cap regions of the Muon Spectrometer (MS).
They provide high timing resolution for precise bunch crossing identification, but have a limited spatial resolution of a few centimetres resulting in a limited muon momentum resolution.
Hence, at most half of the muons with transverse momentum between \GeV{10} and \GeV{20} are selected with a nominal trigger threshold of \GeV{20}.
At HL-LHC luminosities, this leads to unacceptably high trigger rates.

In order to reduce the trigger rate, the momentum resolution of the first level trigger system has to be increased to the level of the current Level-2 trigger resolution based on the Monitored Drift Tubes (MDT) tracking chambers~\cite{LoI}.
The most efficient solution with the best resolution is to use the MDT chambers at the first trigger level~\cite{L1_Improvement2013}.

\section{Using MDT Data for the First-Level Muon Trigger}

In order to include the MDT hit information in the first level trigger, a fast MDT read-out has to be implemented.
Furthermore, a fast track reconstruction algorithm with latency below 6~$\mu$s seeded by the region-of-interest information of the trigger chambers has to be developed.
The track reconstruction has to be performed with high efficiency at high background occupancies of up to 14\% expected in the MDT chambers at HL-LHC~\cite{LoI}. 

\begin{figure}[t]
	\centering
	\includegraphics[width=1.0\linewidth]{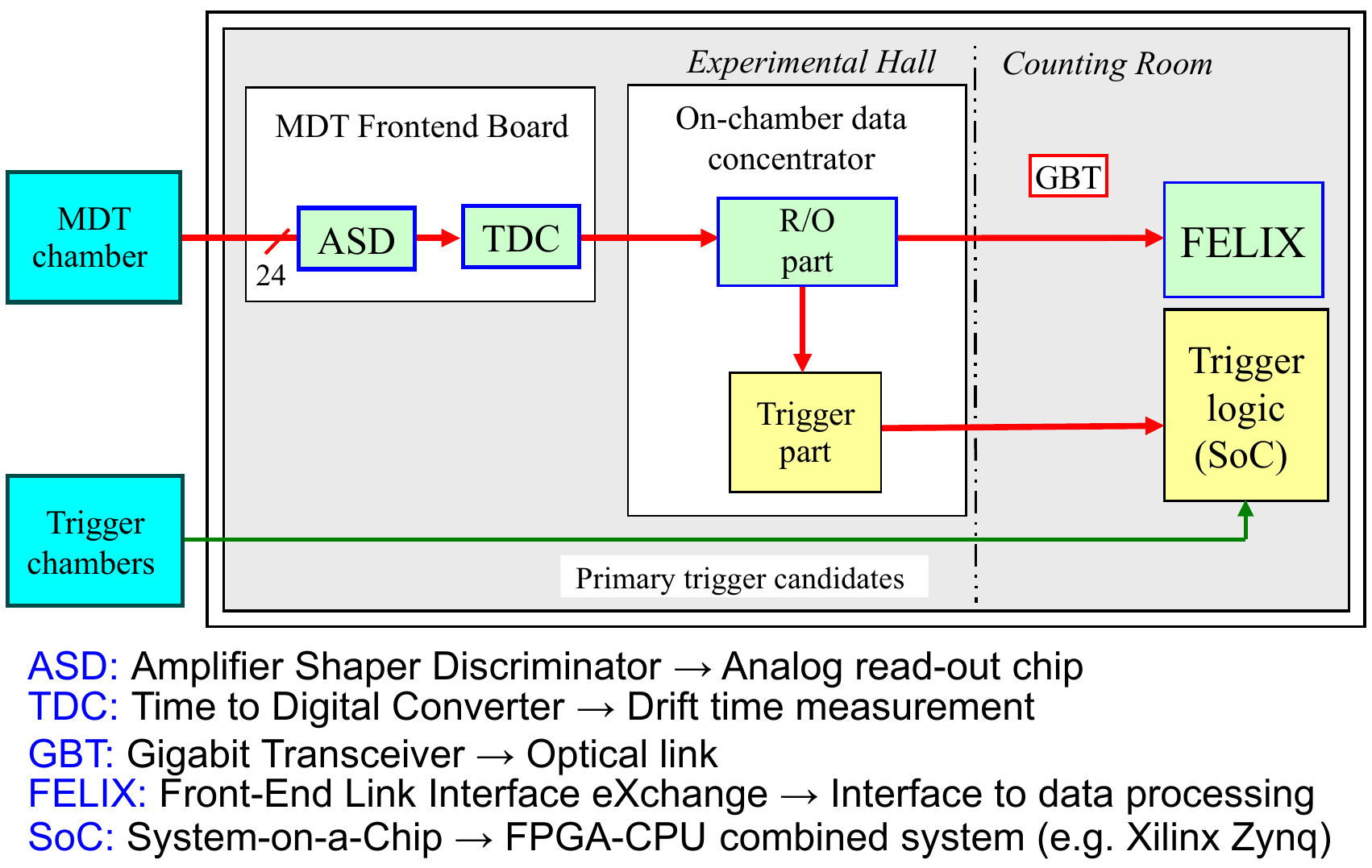}
	\caption{Schematics of the MDT chamber read-out with additional fast read-out path (yellow) for the MDT based first-level muon trigger.}
	\label{fig::technical}
\end{figure}

Figure~\ref{fig::technical} shows the proposed fast MDT read-out scheme.
The digitized drift time measurement of the MDT front-end electronics is sent via the on-chamber data concentrator and a fast read-out path to the trigger logic where fast processors 
(SoCs-``System-on-a-Chip'', FPGA-CPU combinations) are used for hit buffering and track reconstruction.
When the trigger chambers detect a muon candidate, this coarse position information is used for fast track reconstruction within the MDT chambers.

A histogram-based pattern recognition algorithm has been developed for fast track reconstruction~\cite{TIPP} (see \fig{fig::histogram}).
Based on the angle of incidence of the muons known with sufficient precision from the trigger chambers (seed angle), the algorithm resolves the ambiguity in the hit positions from the 
drift radius measurements in the drift tubes and projects the hit coordinates perpendicular to the expected track direction filling a one-dimensional histogram which is scanned for 
peaks corresponding to track candidates.
A linear track fit is applied to the associated hits and the ${\chi}^2$ per degree of freedom of the fit is used as a measure of the fit quality ($\chi^2/N_{dof}<6.6$).
The $\chi^2$-cut leads only to 1.9\% efficiency loss, but reduces the fraction of badly reconstructed tracks\footnote{$|\alpha_{rec}-\alpha_{ref}|>$3~mrad with $\alpha_{rec}$ 
($\alpha_{true}$) the reconstructed (true) track angles.} by 23\%.

In order not to wrongly reject energetic muons, events without a track fulfilling the requirements (less than 10\%) are always accepted and the transverse momentum is determined and 
selected at a higher trigger level~\cite{TIPP}.


\section{Performance Studies}

\begin{figure}[t]
	\centering
	\begin{subfigure}[b]{0.45\linewidth}
		\includegraphics[width=1.1\linewidth]{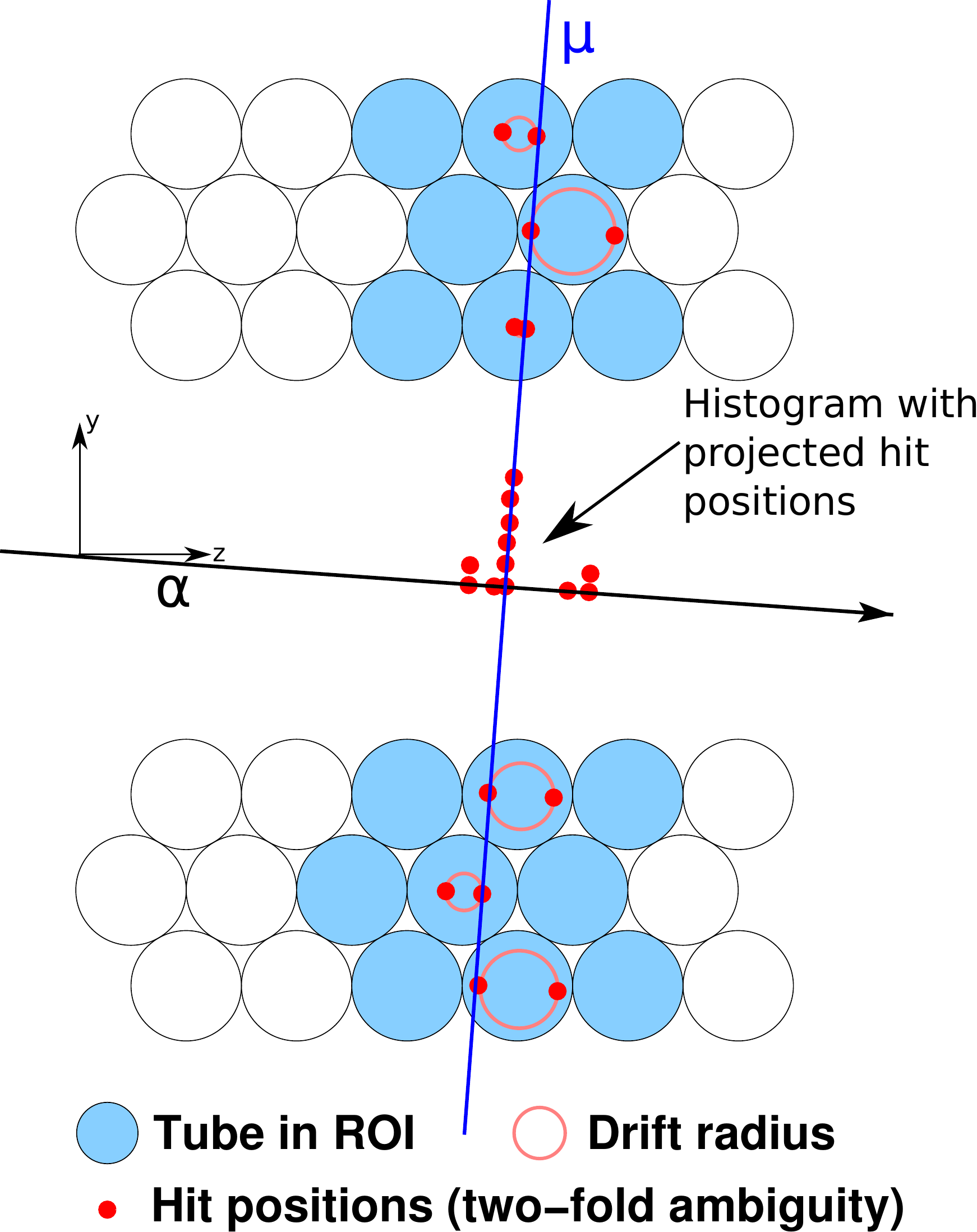}
		\caption{}
		\label{fig::histogram}
	\end{subfigure}
	\qquad
	\begin{subfigure}[b]{0.45\linewidth}
		\includegraphics[width=1.1\linewidth]{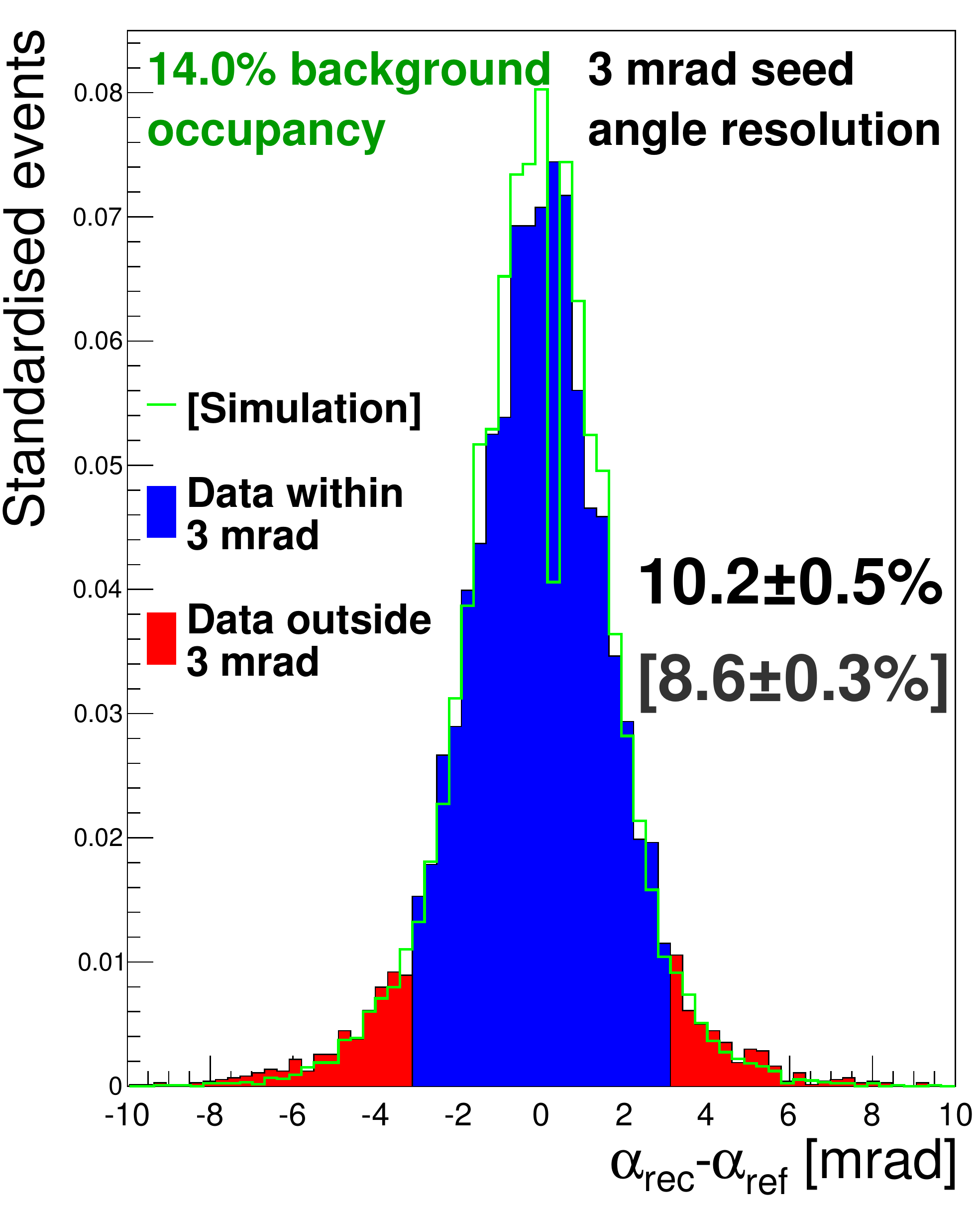}
		\caption{}
		\label{fig::result_irr}
	\end{subfigure}
	\caption{(\subref{fig::histogram}) Illustration of the histogram based pattern recognition method which projects the measured drift radii of the MDT tube layers perpendicular to the expected track direction~\cite{TIPP}.\\
	(\subref{fig::result_irr}) Results of the MDT trigger demonstrator test at the CERN Gamma Irradiation Facility for the maximum background occupancy expected at HL-LHC.
	The black numbers indicate the fraction of low-quality tracks and Geant4 simulation results are given in parentheses.}
	\label{fig::results}
\end{figure}

In order to test the trigger concept and to study the fast track reconstruction algorithm, a hardware demonstrator has been constructed and used for fast reconstruction 
of cosmic muon tracks in an MDT chamber under $\gamma$ irradiation at the CERN Gamma Irradiation Facility (GIF).
The trigger demonstrator comprises two parallel read-out chains for data with high and with reduced time resolution (see \fig{fig::technical}).

The residual distribution of the incidence angle $\alpha_{rec}$ of the fast reconstructed tracks with respect to the offline reconstructed angle $\alpha_{ref}$ is shown in 
\fig{fig::results} and is in agreement with the Geant4~\cite{Geant4_1, Geant4_2} simulation of the demonstrator test set-up.
Due to the preliminary long dead time (1500~ns) of the fast read-out, the fraction of tracks with a small number of hits is enhanced leading to a higher uncertainty in the track fit.
Therefore, the fraction of wrongly reconstructed tracks with $|\alpha_{rec}-\alpha_{ref}|>$3~mrad is 10.2\% at the maximum expected background rate at HL-LHC.
For the implementation in ATLAS, a dead time of 200~ns is envisaged to reduce the fraction to 1\%.


\begin{figure}[t]
	\centering

		\centering
		\includegraphics[width=1.0\linewidth]{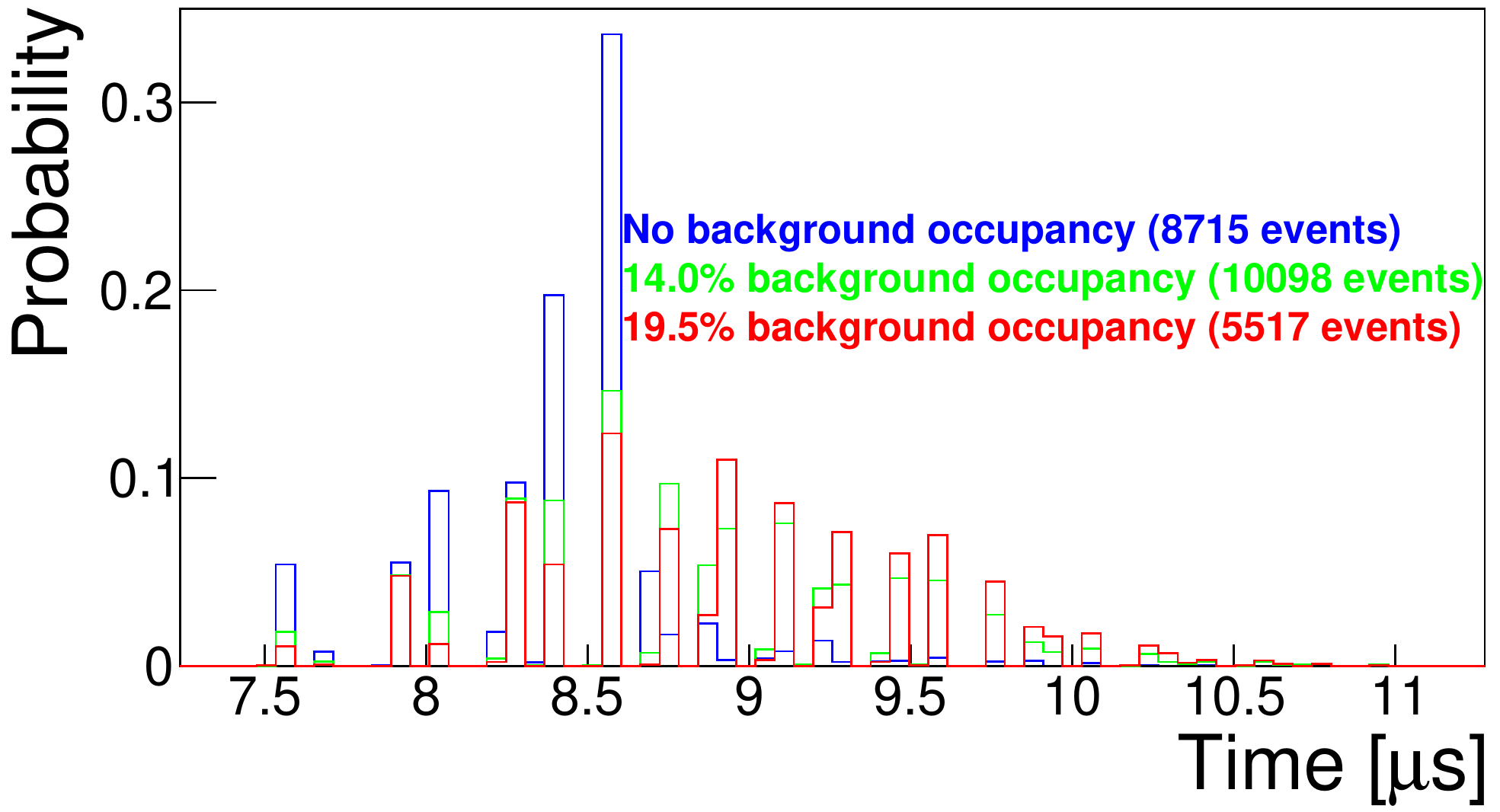}
	\caption{Processing time of the fast track reconstruction algorithm by a 200~MHz CPU for the data taken at the CERN GIF at different background occupancies.}
		\label{fig::timing}
\end{figure}


The fast tracking algorithm has been implemented in assembler language on an ARM Cortex-M4F CPU, commonly used in SoCs.
The processing times of the reconstruction algorithm for the data taken at the GIF are shown in \fig{fig::timing}.
With the highest clock frequency (200~MHz), the track can be reconstructed within 11~$\mu$s.
The use of a more advanced CPU architecture (e.g. ARM Cortex-A9 implemented on Xilinx Zynq SoC) allowing a reduction of the maximum processing time below the expected first-level trigger latency of 6~$\mu$s at HL-LHC.

\section{Summary}

In order to improve the transverse momentum resolution and the acceptance of the ATLAS muon trigger at HL-LHC, the MDT precision tracking chambers will be included in the first level muon trigger with a foreseen latency of 6~$\mu$s.
Performance studies with a hardware implementation of the required new fast read-out of the MDT chambers and of the fast tracking algorithm demonstrate the feasibility of the concept.


\end{document}